\documentclass{aa}
\usepackage{graphicx}

\def\ltsima{$\; \buildrel < \over \sim \;$}
\def\simlt{\lower.5ex\hbox{\ltsima}}
\def\gtsima{$\; \buildrel > \over \sim \;$}
\def\simgt{\lower.5ex\hbox{\gtsima}}

\def\ROSAT{{\it ROSAT} }
\def\SAX{{\it BeppoSAX} }
\def\ASCA{{\it ASCA} }
\def\CHANDRA{{\it Chandra} }
\def\XMM{{\it XMM-Newton} }

\begin{document}
   \title{Disclosing the true nature of the Sy~2 galaxy NGC~3281: \\
one more Compton-thick source }

   \titlerunning{BeppoSAX view of the Sey~2 galaxy 
NGC~3281}
   \authorrunning{C. Vignali \& A. Comastri}

   \author{C. Vignali\inst{1}
\thanks{\emph{Present address:} Department of Astronomy and Astrophysics, 
The Pennsylvania State University, 525 Davey Lab, 
University Park, PA 16802, USA}
          \and
	  A. Comastri\inst{1}
	  }

   \offprints{C. Vignali, \email{chris@astro.psu.edu}}

   \institute{
              Osservatorio Astronomico di Bologna, 
              Via Ranzani 1, I--40127 Bologna, Italy
             }


\abstract{ 
We present the \SAX broad-band X--ray spectrum of the Seyfert~2 galaxy NGC~3281. 
The source high-energy spectrum is 
characterized by the nuclear transmitted component, with an 
absorbing column density of $\approx2$$\times$10$^{24}$ cm$^{-2}$, while the MECS spectrum 
is reflection-dominated, with a prominent \hbox{(EW$\approx$0.5--1.2 keV)} iron K${\alpha}$ emission line. 
The source is detected at only the 5$\sigma$ significance level 
in the LECS band, because of the strong obscuration 
which hampers at low energies the direct view of the active nucleus harbored in NGC~3281. 
\SAX results are consistent with the scenario where NGC~3281 is inclined more than 60$^{\circ}$ 
with respect to the line-of-sight. 
Combining the N$_{\rm H}$ value obtained from the present X--ray analysis with the A$_{\rm V}$ measurement, a 
N$_{\rm H}$/A$_{\rm V}$ about 50 times the Galactic value is derived. 
}
\maketitle

\section{Introduction}

According to the unified models (e.g., Antonucci 1993), Seyfert~2 galaxies are powered by the same 
engine of Seyfert~1s but are seen at medium/high inclination angles with respect to the line-of-sight, 
thus being absorbed by thick matter in the equatorial plane. 
This picture, at least at its first order approximation, has found strong support from many 
observational evidences, including optical spectropolarimetry (e.g., Antonucci \& Miller 1985; Miller \& Goodrich 1990), 
near-infrared spectroscopy (e.g., Goodrich et al. 1994) and X--ray spectroscopy (e.g., Turner et al. 1997a,b; 1998). 
For column densities lower than 10$^{24}$ cm$^{-2}$, the X--rays above a few keV are able to penetrate 
the absorbing medium (likely to be associated to the torus) and to 
be detectable as direct continuum emission (Compton-thin sources). 
If the absorbing column density exceeds $\sim$ 1.5$\times$10$^{24}$ cm$^{-2}$, 
then the nuclear radiation is obscured up to about 
10 keV and the source is called Compton-thick (see Matt et al. 2000 and Matt 2001 for the most recent reviews). 
While in a handful of Compton-thick sources the column density has been directly measured 
(e.g., NGC~6240: Vignati et al. 1999; NGC~4945: Iwasawa et al. 1993; Done et al. 1996; 
Guainazzi et al. 2000; Madejski et al. 2000; 
Circinus: Matt et al. 1999a), in the majority of them the column density is so large to 
totally obscure the nucleus even in the hard X--ray band (e.g., NGC~1068: Matt et al. 1997; 
NGC~7674: Malaguti et al. 1998) or their flux at high energies is too low to 
allow a detailed spectral analysis (Salvati et al. 1997; Maiolino et al. 1998). 

Emission from Compton-thick galaxies can be detected by means of the 
reflection caused by the visible inner surface of the torus and/or via scattering 
by the material responsible for the broad lines observed in polarized light 
(e.g., Antonucci \& Miller 1985; Tran 1995). In the X--ray band, the spectrum of Compton-thick sources 
is generally characterized by an iron K$\alpha$ line with extremely large equivalent width 
(EW$\ga$1 keV) and a reflection-dominated continuum flatter than the intrinsic 
spectrum (e.g., Maiolino et al. 1998). 
In the last few years, thanks to the high capabilities of observing Compton-thick sources 
with broad-band instruments like those onboard \ASCA (e.g., Matt et al. 1996a; Collinge \& Brandt 2000), 
{\it RXTE} (e.g., Madejski et al. 2000), 
\SAX (e.g., Malaguti et al. 1998; Matt et al. 1999a; Cappi et al. 1999; 
Vignati et al. 1999; Guainazzi et al. 2000; Franceschini et al. 2000; 
Iwasawa et al. 2001) and, more recently, \XMM (Guainazzi et al., in preparation), 
much has be learned about their X--ray properties. 
However, there are still several astrophysical reasons why Compton-thick sources deserve to be studied 
in detail. 
About 50\% of the active galactic nuclei (AGNs) in the local Universe are obscured by 
Compton-thick matter (e.g., Maiolino et al. 1998; Risaliti et al. 1999). 
Even though their contribution to the X--ray background (XRB) is not energetically dominant 
(unless the number of such objects is highly underestimated), 
they may constitute however an important ingredient for the IR and the sub-mm backgrounds, where most of the 
absorbed radiation is re-emitted by dust (Fabian \& Iwasawa 1999; Brusa et al. 2001). 
Furthermore, 
Compton-thick sources allow to study those spectral components that otherwise would be 
dominated by the nuclear continuum emission (Matt et al. 1996b; Bianchi et al. 2001). 

In this regard, NGC~3281 is an extremely interesting object. 
\ASCA observation, in a relatively short exposure time ($\approx15$ ks), revealed a complex X--ray spectrum with 
a prominent iron line (equivalent width of $\sim$ 480$^{+770}_{-230}$ eV) 
and a large absorbing column density ($N_{\rm H}$$\simeq$7.1$\pm{1.2}$ $\times$ 10$^{23}$ cm$^{-2}$, 
Simpson 1998). 
The extinction to the nucleus inferred from the near-infrared studies, $A_{\rm V}$=22$\pm{11}$ mag, 
combined with the $N_{\rm H}$ measurement derived from the X--ray analysis, gives a $N_{\rm H}$/$A_{\rm V}$ ratio 
which is more than one order of magnitude larger than the Galactic value 
($N_{\rm H}$/$A_{\rm V}$=1.9$\times$10$^{21}$ cm$^{-2}$ mag$^{-1}$, 
assuming $A_{\rm V}$=3.1\ E(B$-$V), Bohlin et al. 1978). 
Simpson (1998) has explained this finding as due to a dense cloud in the foreground of both 
the X--ray and the infrared emitting regions that obscures the whole X--ray source but only a 
fraction of the much larger IR region. 
There is however increasing evidence from local AGNs 
(Risaliti et al. 2000; Maiolino et al. 2001a; Severgnini et al. 2001) and 
optical/\hbox{X--ray} selected quasars (Risaliti et al. 2001; Fiore et al. 2000; Comastri et al. 2001) that 
the $N_{\rm H}$/$A_{\rm V}$ ratio can be much higher than the Galactic value. 
This ratio may also affect the fraction of nuclear radiation reprocessed and re-emitted by dust 
at far-infrared and sub-millimeter wavelengths.

\section{Observation and data reduction}

The Italian-Dutch satellite \SAX (Boella et al. 1997a) carries 
four co-aligned Narrow-Field Instruments (hereafter NFI), two of which 
are gas scintillation proportional counters with imaging capabilities: 
the Low Energy Concentrator Spectrometer (LECS: 0.1--10 keV, Parmar et 
al. 1997) and the Medium Energy Concentrator Spectrometer (MECS: 1.3--10 
keV, Boella et al. 1997b). The remaining two instruments are the High 
Pressure Gas Scintillation Proportional Counter (HPGSPC: 4--120 keV, Manzo et al. 1997), 
which was switched off during the observation, 
and the Phoswich Detector System (PDS: \hbox{13--200} keV, Frontera et al. 1997). 

NGC~3281 was observed by \SAX from May 20 to May 23 2000. 
Standard reduction techniques and screening criteria were applied in 
order to produce useful linearized and equalized event files. 
The resulting net exposure times are about 71.3~ks, 13.5~ks and 35~ks 
for MECS, LECS and PDS, respectively. 
Spectra from MECS and LECS were extracted from circular regions 
of radius 4$\arcmin$ around the source centroid, while 
background spectra were extracted from both blank-sky event files 
and from source-free regions in the target field-of-view, with no apparent 
difference between the two spectra. 
The background subtraction in PDS light
curves and spectra was performed by plain subtraction of the 
``off-source'' from the ``on-source'' products. The systematic 
uncertainties of this method are lower than $\simeq0.03$~s$^{-1}$ in 
the full PDS sensitive energy bandpass (Guainazzi \& Matteuzzi 1997). 
The background-subtracted count rates are \hbox{2.19$\pm{0.06}$ $\times$ 10$^{-2}$}, 
\hbox{3.36$\pm{0.63}$ $\times$ 10$^{-3}$} and \hbox{0.73$\pm{0.03}$} counts per second 
for MECS (1.3--10 keV band), LECS (0.5--4.5 keV) and PDS (13--100 keV), 
respectively. 

The choice of the extracting regions is such to ensure that the groups of 
galaxies populating NGC~3281 field-of-view (Ferguson \& Sandage 1990) and apparent in 
MECS image have 
a very poor contribution (if any) to the source spectrum in the E$<$10 keV energy range. 
The PDS band, owing to the soft, presumably thermal spectra characterizing these 
groups of galaxies, is totally unaffected. 
A flat-spectrum radio-loud quasar is present in PDS field-of-view 
(PKS~1030$-$357). The extrapolation of its radio flux to the X--ray band 
by means of a power law well matches the {\it ROSAT} All Sky Survey (RASS) detection and 
gives a very poor contribution, of the order of a few 10$^{-3}$, to the 20--100 keV flux. 
No known bright source is present in the PDS field-of-view. 
The probability of a serendipitous source with a flux equal or larger 
than NGC~3281 is $\sim$ 3.9$\times$10$^{-3}$ (assuming the 
\ASCA \hbox{2--10} keV LogN-LogS from Cagnoni et al. 1998). \\
No significant flux variability has been revealed either in the MECS or in the PDS light curve. 

The spectral analysis was carried out through the {\sc XSPEC} program 
(version 11, Arnaud 1996). Errors are quoted at 90 \% confidence level 
for one interesting parameter ($\Delta\chi^{2}$=2.71, Avni 1976), and 
energies are reported in the source rest frame. 
The Anders \& Grevesse (1989) abundances have been used throughout the paper, as well as 
a cosmology with $H_{0}$=70 km s$^{-1}$ Mpc$^{-1}$ and $q_{0}$=0.5 (unless otherwise stated).

\section{Spectral analysis}

NGC~3281 is detected in the LECS at only the $\approx5$$\sigma$ significance level, the cause being the short 
exposure time in the low-energy spectrometer (about 13.5 ks) and presumably 
the large absorption obscuring the source, which previous \ROSAT and {\it Einstein} IPC (Fabbiano et al. 1992) 
observations have already suggested. 
Due to the low quality of the data in the LECS band, in the following only the MECS and PDS data will be 
presented. 
After an  introduction to the broad-band X--ray spectrum of NGC~3281 (Sect.~3.1) 
using the \hbox{1.3--100 keV} MECS$+$PDS data, we will focus on the \hbox{3--100 keV} 
X--ray properties (Sect.~3.2). 
Then the results will be discussed and put into a physically more exhaustive scenario (Sect.~4). 

\subsection{An introduction to the source spectral complexity:\\ 
the 1.3--100 keV spectrum}

To allow for differences in the absolute flux calibration of the MECS and PDS detectors, 
the normalizations of the two instruments were allowed to vary within the fiducial values 
reported by Fiore et al. (1999). 
A column density of 6.42$\times$10$^{20}$ cm$^{-2}$, due to absorption by the Galaxy 
(Dickey \& Lockman 1990), is included in all of the models. 
\begin{figure}
\resizebox{\hsize}{!}{\includegraphics[angle=-90]{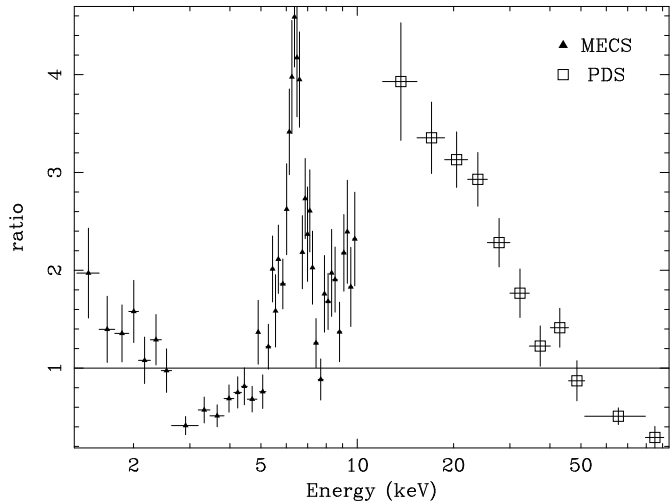}}
\caption{Residuals of the MECS$+$PDS 1.3--100 keV spectrum for a model 
consisting of a single power law with $\Gamma$$\sim$0.3 and Galactic absorption. 
Spectral complexity in the 
soft X--ray band is clearly present, as well as a strong Fe K$\alpha$ emission 
line at about 6.4 keV and an upturn of the spectrum in the PDS band. }
\label{fig1} 
\end{figure}
In Fig.~1 the residuals obtained by fitting the \hbox{1.3--100 keV} spectrum with a 
power-law model plus Galactic absorption are shown. 
The fit ($\chi^{2}$=764 with 96 degrees 
of freedom) provides an extremely flat 
slope ($\Gamma$$\simeq$0.3) and clearly illustrates the complex spectral 
structure below 10 keV and the sharp ``rise and fall'' of the spectrum at 
energies above $\approx8$ keV. 
\begin{figure}
\resizebox{\hsize}{!}{\includegraphics[angle=-90]{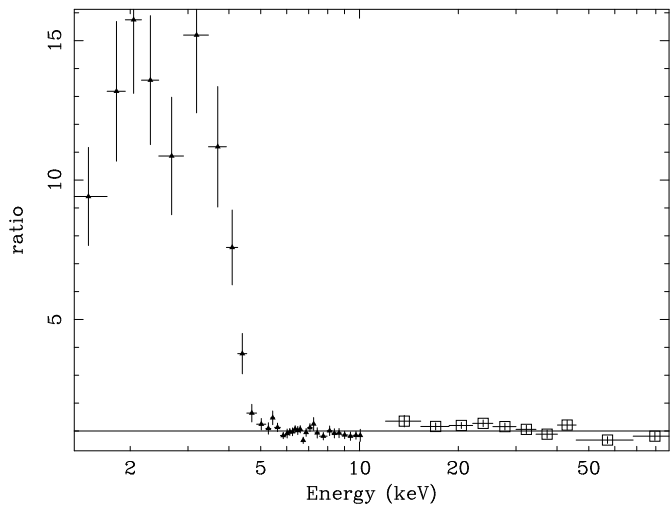}}
\caption{The residuals obtained assuming a 
transmission plus iron K$\alpha$ emission line model. }
\label{fig2} 
\end{figure}
\begin{figure}
\resizebox{\hsize}{!}{\includegraphics[angle=-90]{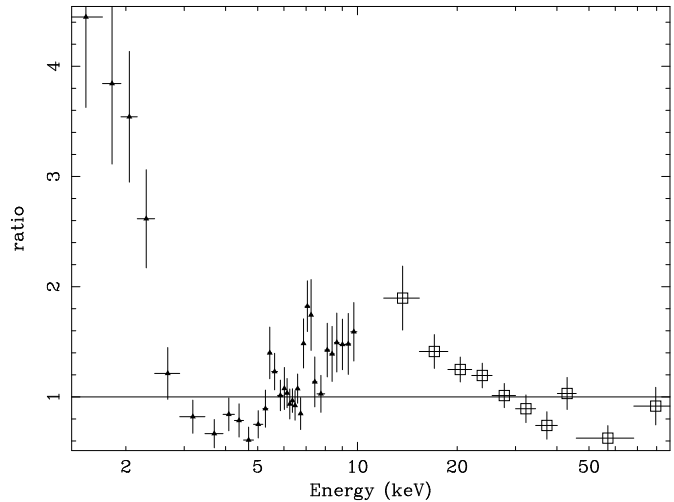}}
\caption{The residuals obtained assuming a 
reflection plus iron K$\alpha$ emission line model. }
\label{fig3} 
\end{figure}
The shape of the residuals suggests the presence of a nuclear, 
absorbed component, which is likely to dominate the emission in the PDS band. 
We therefore tried to fit the data with an absorbed power law model (transmission model) 
plus a narrow iron K$\alpha$ emission line. 
The large excess of counts below $\approx5$ keV (Fig.~2) strongly indicates that this is only 
a rough parameterization of the spectrum of NGC~3281 ($\chi^{2}$/dof=311/93), and 
that the emission in the MECS band has probably a different origin. 

Therefore, owing to the flat spectrum and the presence of a strong Fe K$\alpha$ emission line, 
we tried to fit the overall broad-band spectrum with a pure-reflection model ({\sc PEXRAV} 
in {\sc XSPEC}, Magdziarz \& Zdziarski 1995). 
Although this model provides a statistically significant improvement 
in the spectral fit with respect to the transmission model 
($\Delta\chi^{2}$$\simeq$74 with the same number of dof), 
corresponding to an improvement $>$99.99\% according to the F-test), 
it still represents a bad parameterization of NGC~3281 X--ray emission. 
Indeed, large residuals appear evident (Fig.~3), both in the MECS 
and in the PDS band. 
An additional absorbed power-law component is likely to 
be responsible for most of the X--ray emission in the PDS band (see Sect.~3.2). 

However, neither the transmission nor the reflection model (or a combination of the two) 
are able to reproduce the soft (E$\la$3 keV) 
\hbox{X--ray} spectrum of NGC~3281. 
The deficit of counts in the \hbox{2--4 keV} energy range, coupled with the excess of counts below 2 keV (Fig.~1), 
indicates that a more complex model is required. 
Thermal emission and/or absorption plus emission from 
a hot, photo-ionized medium could be relevant for NGC~3281, as well as for Mrk~3, 
a Compton-thick galaxy which shows a similar spectral behavior (Iwasawa 1995; Cappi et al. 1999). 
A soft X--ray emission was indeed revealed by \ASCA (Simpson 1998) thanks to 
the low-energy spectral coverage down to $\sim$ 0.5 keV provided by the SIS instrument. 
Unfortunately, the nature of this component cannot actually be addressed 
by the present LECS data due to the source low counting statistics in this instrument. 
It should be noted that soft X--ray components are not uncommon in obscured AGNs, as recently 
observed in Compton-thick sources (Guainazzi et al. 1999) or derived from the X--ray analysis of a 
large sample of hard X--ray selected AGNs (Vignali et al. 2001 and references therein). 

In the following we will discuss only the \hbox{3--100 keV} spectrum of NGC~3281, 
demanding the study of the \hbox{E$<$3 keV} spectral complexity to other 
better-suited \hbox{X--ray} instruments, as those onboard \CHANDRA and {\it XMM-Newton}.

\subsection{The 3--100 keV spectrum}

Supported by the broad-band spectral analysis presented in Sect.~3.1, 
a power-law component was added to the transmission plus iron line model 
to mitigate the residuals in the MECS band below $\approx5$ keV (model {\bf (a)} in Table~1). 
While the nuclear transmitted component is characterized by a slope ($\Gamma_{\rm H}$=2.07$^{+0.32}_{-0.27}$) which is 
consistent, within the errors, with that typical of AGNs (e.g., Nandra \& Pounds 1994; Piro et al. 2000), 
the extremely flat 
power law below 10 keV ($\Gamma_{\rm S}$=0.78$^{+0.59}_{-0.27}$) and the prominent iron line 
prompted us to fit the overall broad-band spectrum with the following transmission plus reflection model ({\bf (b)} 
in Table~1) 
\begin{eqnarray}
F(E) & = & [ A_{\rm tr} Tr(E,N_{\rm H},\Gamma_{\rm H}) +  A_{\rm refl} R(E,\Gamma_{\rm H},E_{\rm c})  \cr
~ & ~ & + {\sl K\alpha~emission~line}]  e^{-N_{\rm H,gal}\sigma_{\rm ph}} 
\end{eqnarray}
where $Tr(E,N_{\rm H},\Gamma_{\rm H})$ is the power-law component (with slope $\Gamma_{\rm H}$) 
transmitted through the column density $N_{\rm H}$, 
$R(E,\Gamma_{\rm H},E_{\rm c})$ is the pure-reflection spectrum obtained using {\sc PEXRAV} in {\sc XSPEC} 
with e-folding energy E$_{\rm c}$, A$_{\rm tr}$ and A$_{\rm refl}$ 
are the normalizations of the transmitted and reflected component, respectively, 
and $N_{\rm H,gal}$ is the absorption due to the Galaxy. 
\begin{table*}[!ht]
\centering
\caption[]{MECS$+$PDS: 3--100 keV spectral fits}
\begin{tabular}{lcccccccc}
\hline
{\sf Model} & $\Gamma_{\rm S}$ & $\Gamma_{\rm H}$ & $N_{\rm H}$ & A$_{\rm tr}$ & A$_{\rm refl}$ &
E$_{K\alpha}$ & EW$_{K\alpha}$ & $\chi^{2}$/dof \\
 & & & ($\times$10$^{24}$ cm$^{-2}$) & ($\times$10$^{-2}$) & ($\times$10$^{-3}$) & (keV) & (eV) & \\
\hline
{\bf (a)} & 0.78$^{+0.59}_{-0.27}$ & 2.07$^{+0.32}_{-0.27}$ & 1.51$^{+0.26}_{-0.24}$ & 
3.42$^{+5.91}_{-2.00}$ & & 6.36$^{+0.06}_{-0.07}$ & 2340$^{+750}_{-710}$ & 78.9/65 \\
{\bf (b)} & & 1.95$\pm{0.18}$ & 1.51$^{+0.20}_{-0.19}$ & 2.24$^{+2.03}_{-1.13}$ & 7.41$^{+2.60}_{-2.04}$ & 
6.37$\pm{0.07}$ & 1180$^{+400}_{-361}$ & 69.7/66 \\
{\bf (c)} & & 1.98$^{+0.04}_{-0.21}$ & 1.96$^{+0.19}_{-0.05}$ & 5.26$^{+3.29}_{-0.40}$ & 6.84$^{+0.96}_{-2.04}$ & 
6.52$\pm{0.06}$ & 526$^{+128}_{-144}$ & 83.3/66 \\
\hline
\end{tabular}
\label{tab1}
\end{table*}
The reflection is required at a high-significant level ($>$99.95\%) according to the F-test. 
The best-fit photon index, $\Gamma$=1.95$\pm{0.18}$, is consistent with 
the values reported for Seyfert~1 galaxies, 
thus confirming one of the basic expectations of 
the Unified Models, i.e. the same engine powering both Sey~1 and Sey~2 galaxies. 
The column density, $N_{\rm H}$$\simeq$1.51$\pm{0.20}$ $\times$ 10$^{24}$ cm$^{-2}$ (Fig.~4), indicates that 
NGC~3281 is likely to be a ``border-line object'' between the Compton-thin and the Compton-thick sources. 
The angle subtended by the reflecting matter to the nucleus, given by the 
relative normalization of the reflected versus the direct component (both reported in Table~1), is 
$\approx$ 0.33$\times$2$\pi$. 
The iron K$\alpha$ line rest-frame energy, \hbox{6.37$\pm{0.07}$ keV}, 
is consistent with a neutral or mildly ionized origin, and 
its equivalent width (EW$\approx$1.2 keV), 
although not so extreme for a candidate Compton-thick galaxy, is consistent with the amount of reflection derived 
from the spectral fit. 
This model provides a good fit to the broad-band data (Fig.~5) and the residuals do not indicate 
evident additional spectral components. 
Small residuals at $\approx7$ keV were fitted with an additional emission line, which could be associated to 
either the ionized H-like iron line arising from ionized matter 
(the ``warm mirror'' observed in several Compton-thick galaxies; e.g., Matt et al. 1999, 2000) or to the 7.06 keV 
iron K$\beta$ emission line revealed in a few Compton-thick sources (e.g., Matt et al. 1996a; Malaguti et al. 1998). 
However, this feature is not statistically significant, its EW upper limit being of 610 eV. 

We also tried a different scenario, in which the reflecting matter is partially covered. This would represent 
a situation in which the line-of-sight towards the nucleus is heavily blocked due to absorption, but the optical depth diminishes 
at higher torus latitudes, thus allowing transmission of some of the radiation reflected by the far side of the torus. 
With the present \SAX data, no apparent difference between an unabsorbed reflection and an absorbed one was found 
in either the X--ray spectral fits or in the derived parameters. 
\begin{figure}
\resizebox{\hsize}{!}{\includegraphics[angle=-90]{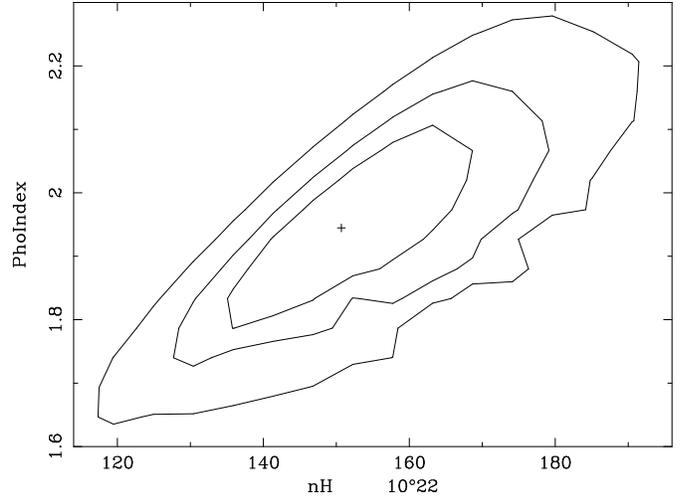}}
\caption{\SAX 68, 90 and 99 \% confidence contours relative to the power-law slope $\Gamma$ vs. the 
column density $N_{\rm H}$ (model {\bf (b)} in Table~1).}
\label{fig4} 
\end{figure}
\begin{figure}
\resizebox{\hsize}{!}{\includegraphics[angle=-90]{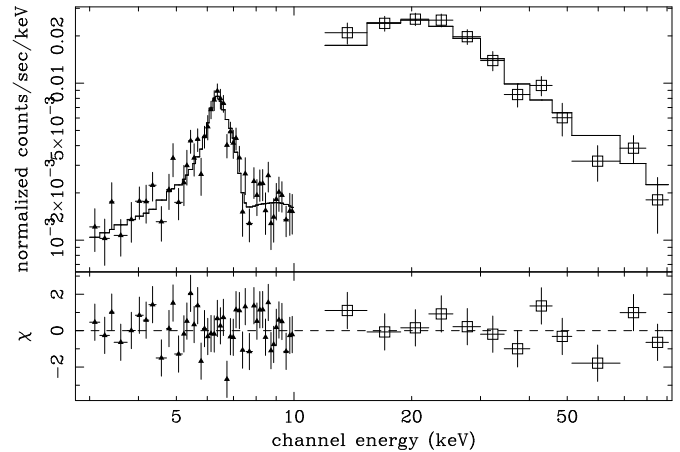}}
\caption{\SAX MECS$+$PDS 3--100 keV spectrum of NGC~3281 (model {\bf (b)} in Table~1). }
\label{fig5} 
\end{figure}

Column densities of the order of 10$^{24}$ cm$^{-2}$ imply that the photons which eventually emerge 
may have been suffered one or more Compton scatterings. 
Therefore we adopted the model of Matt et al. (1999b), based on Monte-Carlo simulations, 
which properly treats the photon transfer in very high-column density matter, including the Compton down-scattering 
and the Klein-Nishina decline, and assuming a spherical geometry for the distribution of the matter 
(model {\bf (c)} in Table~1). 
By this model the column density increases ($N_{\rm H}$=1.96$^{+0.19}_{-0.05}$ $\times$ 10$^{24}$ cm$^{-2}$), while 
both the solid angle subtended by the reflecting matter to the nucleus ($\approx$ 0.13$\times$2\/$\pi$) and the 
iron K$\alpha$ line equivalent width ($\approx500$ eV) decrease. This is due to the increased strength of the nuclear 
transmitted component with respect to the reflected one, as shown in Table~1. 
Even though the fit obtained with such a model is statistically worse ($\chi^{2}$=83.3/66), 
it must be considered physically more correct than those described above, due to the particular treatment of the 
photon transfer required in this heavily absorbed environment. 

Only a lower limit ($\approx60$~keV) was derived for the high-energy cut-off. 
The observed \hbox{2--10 keV} and \hbox{20--100 keV} fluxes of NGC~3281 
derived from the \hbox{best-fit} model are 
\hbox{$\approx$ 2.9$\times$10$^{-12}$} and 
\hbox{7.2$\times$10$^{-11}$ erg cm$^{-2}$ s$^{-1}$}, respectively. 
These fluxes correspond to \hbox{2--10 keV} and \hbox{20--100 keV} intrinsic luminosities of 
\hbox{$\approx$ 1.6$\times$10$^{43}$} and \hbox{2.0$\times$10$^{43}$ erg s$^{-1}$}, respectively 
($\approx$ \hbox{3.1$\times$10$^{43}$} and \hbox{3.9$\times$10$^{43}$ erg s$^{-1}$} adopting 
\hbox{H$_{0}$=50 km s$^{-1}$ Mpc$^{-1}$} and q$_{0}$=0). 
The 2--10 keV flux obtained with \ASCA data in January 1996 observation (Simpson 1998) 
is fully consistent with that derived by {\it BeppoSAX}.

\section{Discussion}

Thanks to its broad-band capabilities, \SAX has been able to reveal the 
X--ray spectrum of the Sey~2 galaxy NGC~3281 being more complex than 
previously derived from \ASCA observation (due to its limited bandpass). 
Below 10 keV, the typical imprints of the reflected continuum 
(i.e., a flat slope and a moderately 
strong iron K$\alpha$ emission line) do clearly appear, and the presence of 
a heavily absorbed component is suggested. 
The reflecting matter appears to be neutral or only marginally ionized, 
since there is no clear evidence of ionized iron features. 

In the PDS band the nuclear, heavily absorbed continuum (N$_{\rm H}$$\simeq$2$\times$10$^{24}$ cm$^{-2}$) 
is able to emerge and to reveal the true nature of NGC~3281, being a Compton-thick source. 
Due to the broad-band spectral coverage, both the reflected and the transmitted components 
have been constrained in a comprehensive picture in which also Compton down-scattering has been taken into 
account. According to this scenario, the strong absorption is due to transmission through the 
rim of the torus, while the unabsorbed (or mildly absorbed) reflection component, directly viewed, is 
due to reprocessing from the torus inner surface. 
This picture agrees with previous findings for Compton-thick sources (e.g., Matt et al. 2000). 

The equivalent width of the iron K$\alpha$ line, coupled to the absorption column density, allows 
to constrain the inclination angle of NGC~3281 with respect to the line-of-sight to be higher than 
60$^{\circ}$ (see Fig.~3 in Ghisellini et al. 1994), assuming typical values for the 
half-opening angle of the torus and its location in the equatorial plane of the galaxy. 
A similar value for the inclination angle was suggested by Storchi-Bergmann (1992) from optical 
observations of this galaxy, in particular from the conical morphology of the high-excitation gas revealed by 
[O~III]$\lambda5007$ emission. 
It is worth noting that the conical morphology, even though not peculiar of Compton-thick objects, 
is expected if large amounts of gas hide the nuclear emission, and was seen in many Sey~2 galaxies 
(Wilson \& Tsvetanov 1994). 
Interesting enough, the [O~III] flux (\hbox{$\approx$ 10$^{-12}$ erg cm$^{-2}$ s$^{-1}$}, 
Storchi-Bergmann et al. 1992) is about 
one-third of the 2--10 keV flux derived by {\it BeppoSAX}. The F$_{[O~III]}$/F$_{2-10 keV}$ ratio 
is extremely high if compared to that obtained by Mulchaey et al. (1994) for a sample of Seyfert~2 galaxies, whose average 
observed ratio is $\approx0.02$. 
This provides further support for the hypothesis of 
obscuration of most of the direct X--rays from NGC~3281 below 10 keV. 

The derived $N_{\rm H}$/$A_{\rm V}$ ratio is about 50 times higher than the Galactic value, 
and a factor of $\approx2$ higher than that derived by Simpson (1998) from the analysis of \ASCA data. 
The high $N_{\rm H}$/$A_{\rm V}$ ratio 
was originally explained by Simpson (1998) assuming a dense cloud which obscures the entire X--ray 
source but only partially the more extended infrared emission. 
An alternative explanation has been recently provided by broad-band studies conducted on AGNs. 
There is, in fact, growing evidence 
(e.g., Severgnini et al. 2001; Maiolino et al. 2001a; Risaliti et al. 2001; Vignali 2001) that such 
high values of $N_{\rm H}$/$A_{\rm V}$ are not exceptions, 
but seem to be a common property of many AGNs at low and high redshift 
along three orders of magnitude in N$_{\rm H}$ (from 10$^{21}$ up to \hbox{10$^{24}$ cm$^{-2}$}, 
Guainazzi et al. 2001). 
Maiolino and collaborators (2001b) have tentatively explained this result 
assuming larger grain sizes and, as a consequence, a flatter extinction curve. 
On the contrary, this anti-correlation between X--ray absorption and optical reddening 
seems not to ubiquitously apply to Compton-thick sources (Guainazzi et al. 2001). 
This result has also cosmological implications. 
Indeed, if 
a significant fraction of the X--ray background is not ascribed only to 
X--ray unabsorbed, broad-line quasars (e.g., Schmidt et al. 1998; Lehmann et al. 2001), 
but actually also to broad-line, dust-free 
X--ray obscured quasars (e.g., Comastri et al. 2001; Fiore et al. 2001), 
then the contribution to the 
IR background of the sources making the hard XRB would be smaller than previously stated 
(e.g., Fabian \& Iwasawa 1999). 

Even though more studies are required to address the above issues, it is however interesting to note 
that only with a broad-band instrument as the PDS onboard \SAX it is possible to reveal the nuclear X--ray 
component in heavily absorbed objects, and to constrain the column density in a accurate way. 
It should be noted, however, that all three models give a fairly good representation of the 
3--100 keV spectrum.

\section{Conclusions}

\SAX observation of the Seyfert~2 galaxy NGC~3281 has allowed to reveal, for the first time, 
its Compton-thick nature. The E$<$10 keV spectrum is reflection-dominated, with a relatively 
strong (EW$\approx$0.5--1.2 keV) 
iron K$\alpha$ emission line. 
The PDS band reveals the presence of the heavily absorbed nuclear continuum, 
the absorption column density being of the order of 1.5--2$\times$10$^{24}$ cm$^{-2}$. 
This finding also justifies the extremely high $F_{[O~III]}$/$F_{2-10~keV}$ value observed in NGC~3281 when compared 
to the average value for Seyfert~2 galaxies. 
The $N_{\rm H}$/$A_{\rm V}$ ratio is extremely high (about 50 times the Galactic value), thus 
adding NGC~3281 to the list of AGNs characterized by dust properties different 
from the Galactic ones. 

The present \SAX results fit well into the unified models scenarios for the Seyfert galaxies and highlight 
the potentialities of the broad-band X--ray spectroscopy with \SAX to study the buried, heavily absorbed 
Seyferts, as well as to discover new Compton-thick galaxies.

\begin{acknowledgements}

This research has made use of the NASA/IPAC Extragalactic
Database (NED) which is operated by the Jet Propulsion Laboratory,
California Institute of Technology, under contract with the National
Aeronautics and Space Administration. 
CV wishes to thank G. Matt who allows to use his spectral code for 
transmission and Compton scattering, G. Brunetti for useful discussions, and the 
referee for helpful comments.  
The authors gratefully acknowledge support from the Italian Space Agency, 
under the contracts ASI 00/IR/103/AS and ASI-ARS-99-75, 
and from the Ministry for University and Research (MURST) under grant Cofin-00-02-36. 
CV also acknowledges financial support of {\it Chandra} X--ray Center grant G01-2100X. 

\end{acknowledgements}


\begin{thebibliography}{}

\bibitem[]{} 
Anders, E., Grevesse, N. 1989, Geochimica et Cosmochimica Acta, 53, 197

\bibitem[]{}
Antonucci, R.~R.~J. 1993, ARA\&A, 31, 473

\bibitem[]{}
Antonucci, R.~R.~J., Miller, J.~S. 1985, ApJ, 297, 621

\bibitem[]{} 
Arnaud, K.~A. 1996. In ``Astronomical Data Analysis Software and 
Systems V'', ed. G. Jacoby, \& J. Barnes, ASP Conf. Series, vol.~101, 17

\bibitem[]{} 
Avni, Y. 1976, ApJ, 210, 642

\bibitem[]{}
Bianchi, S., Matt, G., Iwasawa, K. 2001, MNRAS, 322, 669

\bibitem[]{} 
Boella, G., Butler, R.~C., Perola, G.~C., et al. 1997a, A\&AS, 122, 299

\bibitem[]{} 
Boella, G., Chiappetti, L., Conti, G., et al. 1997b, A\&AS, 122, 327

\bibitem[]{} 
Bohlin, R.~C., Savage, B.~D., Drake, J.~F. 1978, ApJ, 224, 132

\bibitem[]{} 
Brusa, M., Comastri, A., Vignali, C. 2001. 
In ``Galaxy Clusters and the High Redshift Universe Observed in X--rays'', ed. 
D. Neumann, F. Durret, \& J. Tran Thanh Van, in press (astro-ph/0106014)

\bibitem[]{} 
Cagnoni, I, Della Ceca, R., Maccacaro, T. 1998, ApJ, 493, 5

\bibitem[]{}
Cappi, M., Bassani, L., Comastri, A., et al. 1999, A\&A, 344, 857

\bibitem[]{}
Collinge, M.~J., Brandt, W.~N. 2000, MNRAS, 317, L35

\bibitem[]{}
Comastri, A., Fiore, F., Vignali, C., Matt, G., Perola, G.~C., La Franca, F. 
2001, MNRAS, 327, 781

\bibitem[]{}
Dickey, J.~M., Lockman, F.~J. 1990, ARA\&A, 28, 215

\bibitem[]{} 
Done, C., Madejski, G.~M., Smith, D.~A. 1996, ApJ, 463, L63

\bibitem[]{} 
Fabbiano, G., Kim, D.-W., Trinchieri, G. 1992, ApJS, 80, 531

\bibitem[]{} 
Fabian, A.~C., Iwasawa, K. 1999, MNRAS, 303, L34

\bibitem[]{} 
Ferguson, H.~C., Sandage, A. 1990, AJ, 100, 1

\bibitem[]{} 
Fiore, F., Guainazzi, M., Grandi, P. 1999, Cookbook for \SAX NFI spectral analysis

\bibitem[]{} 
Fiore, F., et al. 2000. In ``X--ray Astronomy '999: Stellar Endpoints, AGN, and the Diffuse Background'', 
ed. G. Malaguti, G.~G.~C. Palumbo, \& N. White, in press (astro-ph/0007118)

\bibitem[]{} 
Fiore, F., Comastri, A., La Franca, F., Vignali, C., Matt, G., Perola, G.~C., and the HELLAS collaboration 2001. 
In Proceedings of the ``ESO/ECF/STSCI Workshop on Deep Fields", ed. S. Cristiani, in press (astro-ph/0102041) 

\bibitem[]{} 
Franceschini, A., Bassani, L., Cappi, M., Granato, G.~L., Malaguti, G., Palazzi, E., Persic, M. 2000, 
A\&A, 353, 910

\bibitem[]{} 
Frontera, F., Costa, E., Piro, L., et al. 1997, A\&AS, 122, 357

\bibitem[]{} 
Ghisellini, G., Haardt, F., Matt., G. 1994, MNRAS, 267, 743

\bibitem[]{} 
Goodrich, R.~W., Veilleux, S., Hill, G.~J. 1994, ApJ, 422, 521

\bibitem[]{} 
Guainazzi, M., Matteuzzi, A. 1997, SDC Technical Report

\bibitem[]{} 
Guainazzi, M., Matt, G., Antonelli, L.~A., et al.  1999, MNRAS, 310, 10

\bibitem[]{} 
Guainazzi, M., Matt, G., Brandt, W.~N., Antonelli, L.~A., Barr, P., Bassani, L. 2000, 
A\&A, 356, 474

\bibitem[]{} 
Guainazzi, M., Fiore, F., Matt, G., Perola, G.~C. 2001, MNRAS, 327, 323

\bibitem[]{} 
Guainazzi, M., et al., in preparation

\bibitem[]{} 
Iwasawa, K. 1995, PhD Thesis

\bibitem[]{} 
Iwasawa, K., Koyama, K., Awaki, H., Kunieda, H., Makishima, K., Tsuru, T., 
Ohashi, T., Nakai, N. 1993, ApJ, 409, 155

\bibitem[]{} 
Iwasawa, K., Matt., G., Fabian, A.~C., Bianchi, S., Brandt, W.~N., Guainazzi, M., 
Murayama, T., Taniguchi, Y. 2001, MNRAS, 326, 119

\bibitem[]{} 
Lehmann, I., Hasinger, G., Schmidt, M., et al. 2001, A\&A, 371, 833

\bibitem[]{} 
Madejski, G., Zycki, P., Done, C., Valinia, A., Blanco, P., 
Rothschild, R., Turek, B. 2000, 535, L87

\bibitem[]{} 
Magdziarz, P., Zdziarski, A.~A. 1995, MNRAS, 273, 837

\bibitem[]{} 
Maiolino, R., Salvati, M., Bassani, L., Dadina, M., Della Ceca, R., 
Matt, G., Risaliti, G., Zamorani, G. 1998, A\&A, 338, 781

\bibitem[]{} 
Maiolino, R., Marconi, A., Salvati, M., et al. 2001a, A\&A, 365, 28

\bibitem[]{} 
Maiolino, R., Marconi, A., Oliva, E. 2001b, A\&A, 365, 37

\bibitem[]{} 
Malaguti, G., Palumbo, G.~G.~C., Cappi, M., et al. 1998, A\&A, 331, 519

\bibitem[]{}
Manzo, G., Giarrusso, S., Santangelo, A., et al. 1997, A\&AS, 122, 341

\bibitem[]{}
Matt, G., Fiore, F., Perola, G.~C., et al. 1996a, MNRAS, 281, L69

\bibitem[]{}
Matt, G., Brandt, W.~N., Fabian, A.~C. 1996b, MNRAS, 280, 823

\bibitem[]{}
Matt, G., Guainazzi, M., Frontera, F., et al. 1997, A\&A, 325, L13

\bibitem[]{}
Matt, G., Guainazzi, M., Maiolino, R., et al. 1999a, A\&A, 341, L39

\bibitem[]{}
Matt, G., Pompilio, F., La Franca, F. 1999b, New Astr., 4, 191

\bibitem[]{}
Matt, G., Fabian, A.~C., Guainazzi, M., Iwasawa, K., Bassani, L., Malaguti, G. 
2000, MNRAS, 318, 173

\bibitem[]{}
Matt, G. 2001. In ``Issues in Unification of AGNs'', R. Maiolino, A. Marconi, \& N. Nagar, 
ASP Conf. Series, in press (astro-ph/0107584)

\bibitem[]{}
Miller, J.~S., Goodrich, R.~W. 1990, ApJ, 355, 456

\bibitem[]{}
Mulchaey, J.~S., Koratkar, A., Ward, M.~J., et al. 1994, ApJ, 436, 586

\bibitem[]{}
Nandra, K., Pounds, K.~A. 1994, MNRAS, 268, 405

\bibitem[]{} 
Parmar, A.~N., Martin, D.~D.~E., Bavdaz, M., et al. 1997, A\&AS, 122, 309

\bibitem[]{} 
Piro, L., et al. 2000, Advances in Space Research, 25, 453 

\bibitem[]{}
Risaliti, G., Maiolino, R., Salvati, M. 1999, ApJ, 522, 157

\bibitem[]{}
Risaliti, G., Gilli, R., Maiolino, R., Salvati, M. 2000, A\&A, 357, 13

\bibitem[]{}
Risaliti, G., Marconi, A., Maiolino, R., Salvati, M., Severgnini, P. 2001, A\&A, 371, 37

\bibitem[]{} 
Salvati, M., Bassani, L., Della Ceca, R., Maiolino, R., Matt, G., Zamorani, G. 
1997, A\&A, 323, L1

\bibitem[]{} 
Schmidt, M., Hasinger, G., Gunn, J., et al. 1998, A\&A, 329, 495

\bibitem[]{}
Severgnini, P., Risaliti, G., Marconi, A., Maiolino, R., Salvati, M. 2001, A\&A, 368, 44

\bibitem[]{} 
Simpson, C. 1998, ApJ, 509, 653

\bibitem[]{} 
Storchi-Bergmann, T., Wilson, A.~S., Baldwin, J.~A. 1992, ApJ, 396, 45

\bibitem[]{}
Tran, H.~D. 1995, ApJ, 440, 597

\bibitem[]{}
Turner, T.~J., George, I.~M., Nandra, K., Mushotzky, R.~F. 1997a, ApJS, 113, 23

\bibitem[]{}
Turner, T.~J., George, I.~M., Nandra, K., Mushotzky, R.~F. 1997b, ApJ, 488, 164

\bibitem[]{}
Turner, T.~J., George, I.~M., Nandra, K., Mushotzky, R.~F. 1998, ApJ, 493, 91

\bibitem[]{} 
Vignali, C. 2001, PhD Thesis, University of Bologna

\bibitem[]{} 
Vignali, C., Comastri, A., Fiore, F., La Franca, F. 2001, A\&A, 370, 900

\bibitem[]{} 
Vignati, P., Molendi, S., Matt., G., et al. 1999, A\&A, 349, L57

\bibitem[]{} 
Wilson, A.~S., Tsvetanov, Z.~I. 1994, AJ, 107, 1227

\end{thebibliography}
\end{document}